\documentclass[aoas,preprint]{imsart}

\RequirePackage[OT1]{fontenc}
\RequirePackage{amsthm,amsmath,amssymb}
\RequirePackage{natbib}
\RequirePackage[colorlinks,citecolor=blue,urlcolor=blue]{hyperref}
\usepackage{bm}

\startlocaldefs
\numberwithin{equation}{section}
\theoremstyle{plain}

\endlocaldefs
\usepackage{graphicx}
\usepackage{caption}

\newtheorem{assumption}{Assumption}
\newtheorem*{theorem*}{Theorem}
\newtheorem{theorem}{Theorem}

\newtheorem{example}{Example}

\newtheorem{corollary}{Corollary}
\newtheorem*{corollary*}{Corollary}

\def\bSig\mathbf{\Sigma}

\def\pr{\textnormal{pr}}   
\def\L{\textnormal{L}}
\def\U{\textnormal{U}}
\def\CI{\textnormal{CI}}

\begin{document}

\begin{frontmatter}
\title{Using Missing Types to Improve Partial Identification with Application to a Study of HIV Prevalence in Malawi}
\runtitle{Using Missing Types to Improve Partial Identification}

\begin{aug}
\author{\fnms{Zhichao} \snm{Jiang}\thanksref{m1}\ead[label=e1]{zhichaoj@princeton.edu}}
\and
\author{\fnms{Peng} \snm{Ding}\thanksref{m2}\ead[label=e2]{pengdingpku@berkeley.edu}}
%

\runauthor{Jiang, Z. and Ding, P.}

\affiliation{Princeton University\thanksmark{m1} and University of California, Berkeley\thanksmark{m2}}

\address{ Department of Politics, Princeton University, New Jersey 08544, USA  \\
\printead{e1}\\
Department of Statistics, University of California, Berkeley, California, USA\\
\printead{e2}
}

\end{aug}

\begin{abstract}
Frequently, empirical studies are plagued with missing data. When the data are missing not at random, the parameter of interest is not identifiable in general. Without additional assumptions, we can derive bounds of the parameters of interest, which, unfortunately, are often too wide to be informative. Therefore, it is of great importance to sharpen these worst-case bounds by exploiting additional information.  Traditional missing data analysis uses only the information of the binary missing data indicator, that is, a certain data point is either missing or not. Nevertheless, real data often provide more information than a binary missing data indicator, and they often record different types of missingness. In a motivating HIV status survey, missing data may be due to the units' unwillingness to respond to the survey items or their hospitalization during the visit, and may also be due to the units' temporarily absence or relocation. It is apparent that some missing types are more likely to be missing not at random, but  other missing types are more likely to be missing at random. We show that making full use of the missing types  results in narrower bounds of the parameters of interest. In a real-life example, we demonstrate substantial improvement of more than 50\% reduction in bound widths for estimating the prevalence of HIV in rural Malawi.   As we illustrate using the HIV study, our strategy is also useful for conducting sensitivity analysis by gradually increasing or decreasing the set of types that are missing at random.
In addition, we propose an easy-to-implement method to construct confidence intervals for partially identified parameters with bounds expressed as the minimums and maximums of finite parameters, which is useful for not only our problem but also  many other problems involving  bounds.
\end{abstract}



\end{frontmatter}

\section{An introduction to missing data and partial identification}

Missing data is a common problem for both experimental and observational studies in social and biomedical sciences.
\citet{rubin1976inference} first clarified the missing at random and missing not at random  mechanisms. 
Intuitively, the missing at random assumption requires the missing data mechanism be independent of the missing values themselves conditional on the observed data, but missing not at random allows for such dependence \citep{rubin1976inference}. For more subtle discussion of these concepts, see \citet{mealli2015clarifying}, \citet{rubin1976inference} and \citet{seaman2013meant}. 
Missing at random is a sufficient condition to justify many missing data methods, including likelihood and Bayesian inference \citep[][]{little2014statistical,yang2016note}, multiple imputation \citep[][]{rubin2004multiple}, and inverse probability weighting and doubly robust estimation \citep[][]{bang2005doubly,kang2007demystifying}.      


Unfortunately, however, the missing at random assumption is untestable and it can be too strong in practice. Without making such assumption, researchers derived bounds of the parameters of interest, considering the  worst-case scenarios of the missing data \citep[][]{ding2014identifiability,horowitz1998censoring,horowitz2000nonparametric,manski2003partial,mattei2014identification}. In fact, this idea had an early root in survey nonresponse, but was abandoned by its inventor \citet{cochran1953sampling} because the  bounds are often too wide to be useful. Recognizing the drawbacks of the extreme bounds, some researchers suggested conducting sensitivity analysis to obtain a range of estimates for the parameters  corresponding to a plausible range of  the sensitivity parameter \citep[][]{copas1997inference,molenberghs2001sensitivity,andrea2001methods,scharfstein1999adjusting,vansteelandt2006ignorance}. Some researchers incorporated expert opinions \citep{scharfstein2004construction} to derive narrower bounds. Other researchers imposed parametric assumptions \citep[][]{miao2016identifiability} or used instrumental variables \citep{ma2003identification,shao2016semiparametric, tang2003analysis} to identify parameters of interest.    


Motivated by a longitudinal survey of HIV prevalence in rural Malawi \citep{arpino2014using}, we propose an alternative approach to improve the inference by exploiting the information about different missing types of the outcomes. For instance, the data have recorded that the units' HIV statuses were missing due to different reasons, i.e., the outcomes have different missing types. Some of them were unwilling to respond, some of them were in hospital, some of them were temporarily absent in the survey, some of them moved to another place to live, and some outcomes were missing due to other reasons. It is evident that some missing types may depend on the HIV status, and other missing types are very likely to be independent of the HIV status. Carefully utilizing the information of the missing types can lead to narrower bounds of the partially identified parameters compared to bounds that use only the binary missing data indicators as in the traditional analysis. Moreover, the HIV status satisfies a natural monotonicity, because a person infected at any given time point must be infected at later time points, whereas a person not infected at any given time point cannot be infected at earlier time points. Therefore, we can further improve the bounds with longitudinal HIV data.
We establish theory to quantify the improvement of bounds in both cross-sectional and longitudinal data, and show that the consequential bounds of the HIV prevalence  are  substantially narrower in theory and in our application.

Although it is straightforward to estimate the bounds, it is challenging to construct confidence intervals for the parameters of interest. Importantly, the estimators of our bounds do not follow asymptotic  normal distributions as required by \citet{imbens2004confidence} and \citet{vansteelandt2006ignorance}, and the bootstrap may lead to invalid asymptotic confidence intervals \citep{andrews2000inconsistency, romano2010inference}. 
We  propose a method to construct confidence intervals for partially identified parameters with upper and lower bounds expressed as the minimums and maximums of finite parameters. Our method  is easy to implement, and is useful for not only our problem but also many other problems involving bound analysis in the missing data and causal inference literature \citep[][]{manski2003partial,cheng2006bounds,mealli2013using,mattei2014identification,yang2016using}.

More practically, our paper offers a novel strategy to conduct sensitivity analysis with respect to the missing data mechanism. Illustrated by the HIV study, we can gradually increase or decrease the set of types that are missing at random, and therefore obtain a sequence of results under assumptions with different restrictions on the missing data mechanism.


\section{Missing data at a single time point}\label{sec::single}

\subsection{Traditional bounds with data missing not at random}\label{sec::traditional-single}

We first consider the case with a single observational time point $t.$ 
Let $Y_t$ be a binary outcome of interest.  For example, in our application, $Y_t$ is the HIV status at a given time $t$, with $Y_t=1$ if infected and $Y_t=0$ otherwise. We focus on a binary outcome, and will comment on general outcomes in Section \ref{sec::discussion}. We condition on covariates implicitly in the theoretical discussion, and all conclusions hold within each stratum of covariates.   Let $S_t$ be the survival status at a given time $t$, with $S_t=1$ if survive and $S_t=0$ otherwise.
 The outcome $Y_t$ is well-defined only  for people alive at time point $t$, i.e., for units with $S_t=0$, we define $Y_t=*$. In addition, $Y_t$  is missing for some alive units.
Let $\tilde{R}_t$ be the missing data indicator, with $\tilde{R}_t=1$ if the outcome is observed, $\tilde{R}_t=0$ if the outcome is missing,  and $\tilde{R}_t=*$ if the unit is dead. 
In many real-world applications, an important quantity of interest is
$$
\pi_t=\pr(Y_t=1\mid S_t=1),
$$ 
which, in the HIV example, is the prevalence of HIV of the alive people at time point $t.$ 

The pattern mixture decomposition \citep{little2014statistical} of the outcome distribution is 
\begin{eqnarray}\label{eq::pattern-mixture}
\nonumber \pi_t &=&  \pr(Y_t=1\mid \tilde{R}_t=1, S_t=1)\pr(\tilde{R}_t=1 \mid S_t=1)\\
 &&+\pr(Y_t=1\mid \tilde{R}_t=0, S_t=1)\pr(\tilde{R}_t=0\mid S_t=1).
\end{eqnarray} 
An advantage of the above decomposition is its transparency for identification analysis. 
For the right-hand side of \eqref{eq::pattern-mixture}, the observed data of survivors allow for identification of $\pr(\tilde{R}_t=1\mid S_t=1)$, $\pr(\tilde{R}_t=0\mid S_t=1)$ and $\pr(Y_t=1\mid \tilde{R}_t=1, S_t=1)$, but do not contain any information about $\pr(Y_t=1\mid \tilde{R}_t=0, S_t=1)$ without further assumptions. Consequently, in general, we can obtain only the bounds of $\pi_t$ by setting $\pr(Y_t=1\mid \tilde{R}_t=0, S_t=1)$ to its extreme values. Because $0 \leq \pr(Y_t=1\mid \tilde{R}_t=0, S_t=1) \leq 1$,  the lower and upper bounds of $\pi_t$ are $ \text{\textnormal{LB}}_t   \leq  \pi_t \leq  \text{\textnormal{UB}}_t$, where
{\small
\begin{eqnarray}
\label{eq:naiveL}\text{\textnormal{LB}}_t &=&  \pr(Y_t=1\mid \tilde{R}_t=1,S_t=1)\pr(\tilde{R}_t=1\mid S_t=1), \\
\label{eq:naiveU}   \text{\textnormal{UB}}_t&=&  \pr(Y_t=1\mid \tilde{R}_t=1, S_t=1)\pr(\tilde{R}_t=1)+\pr(\tilde{R}_t=0\mid S_t=1).
\end{eqnarray}
}

This type of bound analysis, considering the worst case scenarios, has a long history in statistics at least dating back to \citet{cochran1953sampling}'s discussion in survey nonresponse. Similar ideas are extensively explored in econometrics \citep[][]{manski2003partial}. Although the bounds in \eqref{eq:naiveL} and \eqref{eq:naiveU} do not rely on any assumptions about the missing data mechanism, they often correspond to unrealistic extreme cases that all the missing outcomes are $1$ or $0$, and consequently they are too wide to be useful as pointed out by \citet{cochran1953sampling}.  \citet{Rubingreenland2005multiple} echoed this view. 
Therefore, to make useful inference with data missing not at random, we need to exploit more information from the data and background knowledge to sharpen the bounds.

\subsection{Using nonresponse types to sharpen bounds of a single time point}\label{sec::single-types}

The discussion in Section \ref{sec::traditional-single} uses only the binary information of the missing data indicator. In many applications, the data provide additional information about different types of nonresponse. For theoretical discussion, we consider a generic case with a four-valued type-specific missing data indicator. Let $R_t = 1$ if the outcome is observed, $R_t = -1$ if nonresponse is due to reasons related to the missing outcome,  $R_t = 0$ if nonresponse is due to other reasons unrelated to the missing outcome,  and $R_t = *$ if the unit is dead. Real problems, such as the HIV study considered in Section \ref{sec::app}, often record many reasons of nonresponse, but we can collapse the reasons into two categories of nonresponse with $R_t=-1$ or $0$. An example for the first type with $R_t=-1$ is that the unit refuses to answer the survey item about the outcome, and an example of the second type with $R_t=0$ is that the unit moves to another place during the visit. Therefore, the coarsened binary missing data indicator $\tilde{R}_t$ equals $1$ if and only if the type-specific missing data indicator $R_t$ equals $1$, $\tilde{R}_t$ equals $0$ if $R_t=0$ or $-1$, and $\tilde{R}_t$ equals $*$ if $R_t=*$.

The bounds in \eqref{eq:naiveL} and  \eqref{eq:naiveU} do not take into account the difference in nonresponse types. We can improve them, by making full use of the missing data indicator $R_t$ and the following assumption.

\begin{assumption}
\label{asm:pmar}
$\pr(Y_t=1\mid R_t=0, S_t=1) =\pr(Y_t=1 \mid S_t=1)$.
\end{assumption}

Assumption \ref{asm:pmar} states that among survivors, the outcome distribution of the individuals with $R_t=0$ is the same as the outcome distribution of the whole population. According to Bayes' Theorem, Assumption \ref{asm:pmar} is equivalent to $\pr(R_t=0\mid Y_t=1,  S_t=1)=\pr(R_t=0 \mid  S_t=1)$. Therefore, Assumption \ref{asm:pmar} means that the type with $R_t=0$ is {\it missing completely at random}, but it is weaker than the usual missing completely at random assumption.  Recall that in the theoretical discussion, we condition on all the covariates implicitly. With covariates $X_t$ at time $t$, Assumption 1 becomes $\pr(Y_t=1\mid X_t,R_t=0,  S_t=1) =\pr(Y_t=1\mid X_t,  S_t=1)$, meaning that the type with $R_t=0$ is {\it missing at random}.  Therefore, we refer to Assumption \ref{asm:pmar} as the partial missing at random assumption.

 In our motivating example, Assumption \ref{asm:pmar} means that the nonresponse corresponding to $R_t=0$ is due to reasons unrelated to the HIV status. In the data, some individuals' HIV status is missing due to  carelessness in data collection, which is purely random.  Some individuals' HIV status is missing because he/she  moved to another place during the survey, which is also independent of his/her HIV status.  It is plausible to assume that units with these missing types constitute a simple random sample of all the units, or, equivalently, Assumption \ref{asm:pmar} holds.
  In addition to these reasons, there are some other reasons such as ``hospitalization'' and ``refused to answer,'' which are probably  related to the HIV status. As a result, we define these reasons as $R_t=-1$. 
   Not surprisingly, Assumption \ref{asm:pmar} helps to sharpen the bounds of $\pi_t$.

\begin{theorem}
\label{thm:single}
Under Assumption \ref{asm:pmar}, the sharp bounds of $\pi_t$ are $\widetilde{\textnormal{LB}}_t\leq \pi_t\leq  \widetilde{\textnormal{UB}}_t $, where
\begin{eqnarray*}
 \resizebox{1\hsize}{!}{ $\widetilde{\textnormal{LB}}_t=\frac{\pr(Y_t=1, R_t=1\mid S_t=1)}{1-\pr(R_t=0\mid S_t=1)},\quad 
\widetilde{\textnormal{UB}}_t =\frac{\pr(Y_t=1, R_t=1\mid S_t=1)+\pr(R_t=-1\mid S_t=1)}{1-\pr(R_t=0\mid S_t=1)}.$}
\end{eqnarray*}
Moreover,  $[\widetilde{\textnormal{LB}}_t,\widetilde{\textnormal{UB}}_t] \subseteq [\textnormal{LB}_t,\textnormal{UB}_t]$, with $[\widetilde{\textnormal{LB}}_t,\widetilde{\textnormal{UB}}_t] \subset [\textnormal{LB}_t,\textnormal{UB}_t] \subset [0, 1]$ if $\pr(R_t=0) > 0$.
\end{theorem}

Theorem \ref{thm:single} demonstrates that by considering different nonresponse types, we can sharpen the bounds of $\pi_t$ in \eqref{eq:naiveL} and \eqref{eq:naiveU}. The reduction in width of the bounds depends on $\pr(R_t=0 \mid S_t=1)$, the proportion of units with outcomes missing at random.  When $\pr(R_t=0\mid S_t=1)=0$, the bounds are the same as the traditional bounds \eqref{eq:naiveL} and \eqref{eq:naiveU}; when $\pr(R_t=0\mid S_t=1)=1-\pr(R_t=1\mid S_t=1)$, i.e., $\pr(R_t=-1 \mid S_t=1)=0$,  the bounds collapse to a point, and $\pi_t$ is pointly identifiable. A larger value of $\pr(R_t=0\mid S_t=1)$ leads to a larger reduction in the width of the bounds by taking into account  the difference in nonresponse types.
 A practical issue of applying Theorem \ref{thm:single} to our real data is that  we do not know whether some reasons depend on the HIV status or not. We would obtain different bounds if we define different reasons as $R_t=0$. To deal with this issue, we will propose an approach in Section 5.4 to conducting sensitivity analysis based on Theorem \ref{thm:single}.

Assumption \ref{asm:pmar} is weaker than missing at random, because it imposes a partial missing at random assumption that one category of nonresponse is independent of the missing outcome. Assumption \ref{asm:pmar} is similar to the identifying restrictions in pattern mixture model \citep{little1993pattern}, where the parameters for the non-respondents are assumed to be the same as those for the respondents. In the literature,  \citet{harel2009partial} and \citet{little2016conditions} discussed the role of some other partial missing at random assumptions in likelihood and Bayesian inference. We focus on  nonparametric identification and bound analysis without imposing parametric modeling assumptions that are often required by likelihood and Bayesian inferences.

\section{Longitudinal data missing not at random}\label{sec::longitudinal}

\subsection{Improving bounds using nonresponse types with longitudinal data}\label{sec::long-bounds}
When longitudinal data are available, the information across different time points can help improve the bounds, as recognized by \citet{arpino2014using}. For instance, HIV infection is an absorbing state, i.e., a person infected at any given time point must be infected at later time points, whereas a person not infected at any given time point cannot be infected at earlier time points. Mathematically, for  the people alive at both time points $s$ and $t$ with $s<t$, if $Y_t=0$ then $Y_s=0$, and if $Y_s=1$ then $Y_t=1$. We formally state this assumption.

\begin{assumption}
\label{asm:strmon}
For any  $s<t$, if $S_t=S_s=1$, then $Y_s \leq  Y_t$  .
\end{assumption}

Assumption \ref{asm:strmon} imposes monotonicity on the individual level outcomes, which holds naturally for diseases like HIV.  Under monotonicity, for $s<t$, if $Y_{s} =1$, $R_{s} =1$ and $S_t=1$, then  we can infer  $Y_t=1$. Note that we do not set the value of $R_t$ to 1 in this case because the original missing mechanism contains information for the bounds. Other different forms of monotonicity are also used as identification assumptions \citep[e.g.,][]{angrist1996identification,jin2008principal,jiang2016principal}.

Under Assumption \ref{asm:strmon}, we can sharpen the bounds in Theorem \ref{thm:single} with data at multiple time points.

\begin{theorem}
\label{thm:arbitrary}
With data at time points $t-I,\ldots,t,\ldots, t+J$,  under Assumptions \ref{asm:pmar}  and  \ref{asm:strmon},  the sharp bounds of $\pi_t$ are $ \widetilde{\textnormal{LB}}^{I}_t\leq \pi_t \leq \widetilde{\textnormal{UB}}^{J}_t$, where
{\footnotesize
$$
\begin{array}{lll}
&\widetilde{\textnormal{LB}}^{I}_t= \max \Big\{ & \frac{\sum_{i=1}^I \pr(Y_{t-i}=1, R_{t-i}=1,R_{t-i+1} \neq 1, \ldots, R_{t-1}\neq 1,R_t=-1 \mid S_t=1)}{1-\pr(R_t=0 \mid S_t=1)} +\widetilde{\textnormal{LB}}_t,    \\
 && \sum_{i=1}^I \pr(Y_{t-i}=1, R_{t-i}=1,R_{t-i+1}\neq 1, \ldots, R_{t-1}\neq 1 \mid R_t=0, S_t=1) \Big \}, \\
&\widetilde{\textnormal{UB}}^{J}_t= \min \Big\{ & \widetilde{\textnormal{UB}}_t-\frac{\sum_{j=1}^J \pr(Y_{t+j}=0, R_{t+j}=1,R_{t+j-1}\neq 1, \ldots, R_{t+1}\neq 1,R_t=-1, S_{t+j}=1 \mid S_t=1)}{1-\pr(R_t=0 \mid S_t=1)},\\
&&  1- \sum_{j=1}^J \pr(Y_{t+j}=0, R_{t+j}=1,R_{t+j-1}\neq 1, \ldots,  R_{t+1}\neq 1, S_{t+j}=1 \mid R_t=0,S_t=1) \Big \},
\end{array}
$$
}
recalling $\widetilde{\textnormal{LB}}_t$ and  $\widetilde{\textnormal{UB}}_t$ are defined in Theorem \ref{thm:single}.
\end{theorem}
  Because people alive at time point $t$ must be alive at time points before $t$, for $t'<t$, we can observe  $R_{t'}$ for units with $S_t=1$. Thus, we can calculate $\widetilde{\textnormal{LB}}^{I}_t$ from the observed data. Similarly, for units with $S_{t+j}=1$, we observe $(R_{t+j}, R_{t+j-1}, \ldots, R_t)$, and thus we can also calculate $\widetilde{\textnormal{UB}}^{J}_t$ from the observed data.
Note that, to obtain the bounds of $\pi_t$ with longitudinal data, we require Assumption \ref{asm:pmar}  holds  only for time point $t$.
Theorem \ref{thm:arbitrary} shows that the data at time points later than $t$ do not have any information to improve the lower bound of $\pi_t$. Intuitively, this is because the monotonicity assumption is one-sided, i.e., for a positive $j$, we can infer $Y_t=0$ as long as $Y_{t+j}=0$, but we are unsure of $Y_t$ if $Y_{t+j}=1$. Thus, the data at time points later than $t$ can rule out only the possibility that some units' $Y_t$'s take value $1$ but not $0$, and hence do not affect the lower bound.
From a more theoretical view,  with data at time points later than $t$, the unidentifiable term $\pr(Y_t=1 \mid R_t=-1, S_t=1)$ can no longer attain the extreme value $1$ as in Theorem \ref{thm:single}, but it can always attain the extreme value $0$, which keeps the lower bound unchanged. Similar discussion applies to the upper bound.

Comparing Theorems \ref{thm:single} and  \ref{thm:arbitrary}, we can see that the bounds with multiple time points are narrower than the bounds with a single time point. Therefore, collecting data at more time points can always improve the bounds, as long as Assumptions \ref{asm:pmar}  and  \ref{asm:strmon} hold.

In our application,  monotonicity holds automatically for the HIV infection. For other diseases or other types of outcomes, this assumption may be violated. However, without monotonicity, we can still improve the bounds of \citet{arpino2014using} with Assumption \ref{asm:pmar} as shown in Theorem \ref{thm:single}.

\subsection{Testable conditions for Assumptions \ref{asm:pmar}  and  \ref{asm:strmon}}\label{sec::testable-conditions}

Assumptions  \ref{asm:pmar} and \ref{asm:strmon} are crucial for our theory. Therefore, it is important to check their validity empirically.
Requiring the lower bound to be less than or equal to the upper bound in Theorem \ref{thm:arbitrary}, we can obtain testable conditions implied by  Assumptions \ref{asm:pmar} and \ref{asm:strmon}. Moreover, because the bounds in  Theorem \ref{thm:arbitrary} are sharp, these conditions include all testable conditions.  Specifically, $\widetilde{\textnormal{LB}}^{I}_t\leq \widetilde{\textnormal{UB}}^{J}_t$ implies four testable inequalities.
   For simplicity, we give the testable conditions with two or three time points in the main text which are adequate for our empirical application. We can similarly derive the general testable conditions implied by Theorem \ref{thm:arbitrary}, but  to avoid notational complexity we give them in the supplementary material.

\begin{corollary}
\label{thm:con+1}
With data at time points   $t$  and $t+1$, a testable condition  for Assumptions  \ref{asm:pmar} and \ref{asm:strmon} is
$
\widetilde{\textnormal{LB}}_t  \leq 1- \pr(Y_{t+1}=0,R_{t+1}=1, S_{t+1}=1 \mid R_t=0, S_t=1),
$
and it is the only testable condition.
\end{corollary}

The condition in Corollary \ref{thm:con+1} is testable because it depends only on the distribution of the observed data, but does not depend on any missing values. Therefore, the testable condition allows us to falsify Assumptions  \ref{asm:pmar} and \ref{asm:strmon} by the observed data in some scenarios. If the condition is violated, then the data invalidate the fundamental assumptions we make. We give a numerical example to illustrate Corollary \ref{thm:con+1}.

 \begin{example}
 \label{ex:con}
 Suppose that we have data at  time points  $t$  and $t+1$, and for simplicity, all the units are alive at both time points. Thus, we  omit $S_t$ and $S_{t+1}$ in probabilities. Let the  distributions of $Y_{t}$  be $\pr(Y_{t}=1)=2/5$, and assume the following conditional distribution of $R_t$ given $Y_t$: 
\begin{center}
\begin{tabular}{cccc}\hline
$\pr(R_t=r\mid Y_t=y)$ & $R_t=-1$  & $R_t=0$  &$R_t=1$\\ \hline
$Y_t=1$  & $0$  &  $1/4$  & $3/4$\\
$Y_t=0$ & $1/2$  &  $1/2$   &$0$\\ \hline
\end{tabular}
\end{center}
We further assume that the conditional distribution of $(Y_{t+1},R_{t+1})$ on $(Y_{t},R_{t})$ can be decomposed as 
$\pr(Y_{t+1},R_{t+1} \mid Y_{t},R_{t})=\pr(Y_{t+1} \mid Y_{t})\pr(R_{t+1} \mid R_{t})$ with $\pr(Y_{t+1}=1\mid Y_{t}=1)=1$ and $\pr(Y_{t+1}=1\mid Y_{t}=0)=1/6$. Therefore, Assumption 2 holds, and $\pr(Y_{t+1}=1)=1/2$.  The probabilities of the conditional distribution of $R_{t+1}$ on $R_t$ are  
\begin{center}
\begin{tabular}{lccc}\hline
$\pr(R_{t+1}=r_{t+1}\mid R_t=r_t)$ & $R_t=-1$  & $R_t=0$  &$R_t=1$\\ \hline
$R_{t+1}=1$  & $1/2$  &  $1$  & $1/2$\\
$R_{t+1}=0$ & $1/2$  &  $0$   &$1/2$\\
$R_{t+1}=-1$ & $0$ &  $0$   &$0$\\ \hline
\end{tabular}
\end{center}
%
Therefore, the data generating process  violates  Assumption 1.

From the observed data, we can verify that
$
\widetilde{\textnormal{LB}}_t=1/2 >  3/8= 1-\pr(Y_{t+1}=0, R_{t+1}=1 \mid R_t = 0),
$
which violates the condition  in Corollary 1. Therefore, the observed distribution falsifies the conjunction of Assumptions 1 and 2. If we have the prior knowledge that Assumption 2 holds as in the motivating HIV example, then the observed data can falsify Assumption 1. As a result, our bounds are not applicable in this case. 
\end{example}

Corollary \ref{thm:con+1} shows that, although the missing data mechanism in Assumption \ref{asm:pmar} cannot be tested alone, it can be tested when Assumption \ref{asm:strmon} holds a priori. For related discussion on testable conditions in other contexts, please see \citet{balke1997bounds}, \citet{cheng2006bounds} and \citet{kitagawa2015test}.
The following  corollary gives the testable condition with time points $t-1$ and $t$.  

\begin{corollary}
\label{cor::con-1}
With data at time points   $t-1$  and $t$, a testable condition for Assumptions 1 and 2 is
$
\pr(Y_{t-1}=1,R_{t-1}=1 \mid R_t=0, S_t=1)\leq   \widetilde{\textnormal{UB}}_t,
$
and it is the only testable condition.
\end{corollary}
Because our HIV example has three time points, we present theoretical results that are directly applicable.
With three time points $t-1$, $t$ and $t+1$, the sharp lower and upper bounds of $\pi_t$  are  $\widetilde{\textnormal{LB}}^{1}_t$ and $\widetilde{\textnormal{UB}}^{1}_t$. Thus, we have the following corollary.

\begin{corollary}
\label{cor::con+1-1}
With data at time points  $t-1$, $t$  and $t+1$, testable conditions for Assumptions  1 and 2 are
{\footnotesize
$$
\begin{array}{rclcl}
\pr(Y_{t+1}=0,R_{t+1}=1, S_{t+1}=1 \mid R_t=-1, S_t=1)&+&\pr(Y_{t-1}=1,R_{t-1}=1\mid R_t=-1, S_t=1) &\leq& 1,\\
\pr(Y_{t+1}=0,R_{t+1}=1, S_{t+1}=1 \mid R_t=0, S_t=1)&+&\pr(Y_{t-1}=1,R_{t-1}=1\mid R_t=0, S_t=1) &\leq& 1,\\
\pr(Y_{t+1}=0,R_{t+1}=1, S_{t+1}=1 \mid R_t=0, S_t=1)  &+& \frac{\pr(Y_{t-1}=1, R_{t-1}=1, R_t=-1\mid  S_t=1)}{1-\pr(R_t=0\mid S_t=1)}
&\leq& 1-  \widetilde{\textnormal{LB}}_t ,\\
\pr(Y_{t-1}=1,R_{t-1}=1 \mid R_t=0, S_t=1)  &+&\frac{\pr(Y_{t+1}=0, R_{t+1}=1, R_t=-1, S_{t+1}=1\mid S_t=1)}{1-\pr(R_t=0\mid S_t=1)}
 &\leq&  \widetilde{\textnormal{UB}}_t.
\end{array}
$$
} 
\end{corollary}

In practice, we should first check the testable conditions  before reporting the bounds. If the conditions are violated, then the data provide evidence against Assumptions \ref{asm:pmar} and \ref{asm:strmon}. In Section \ref{sec::app}, our real data have observations at three time points and Assumption \ref{asm:strmon} holds by nature of HIV. Fortunately, the data do not contradict any testable conditions.  Therefore, although the data cannot validate Assumption \ref{asm:pmar}, they provide no evidence against it.

\section{Confidence intervals for partially identified parameters}
\label{sec::new-ci}

It is relatively easy to obtain point estimates of the lower and upper bounds by replacing the probabilities by their sample frequencies. To account for sample variability, we need to construct a confidence interval for $\pi_t$. However, because the lower and upper bounds are the minimums or maximums of some parameters, their moment estimators are not asymptotically normal. Thus, we cannot use traditional techniques without further modifications \citep{andrews2000inconsistency, imbens2004confidence, vansteelandt2006ignorance, romano2010inference}. In this section, we propose a method to obtain valid confidence intervals for partially identified parameters of certain forms.

 \subsection{Method and theory conditioning on covariates}
\label{subsec::new-ci1}
We drop the subscript $t$ for notational simplicity. In the previous sections, the parameter of interest, $\pi$, has bounds of the following form:
\begin{eqnarray}
\max\{ \L(1), \ldots, \L(Q) \}  \leq \pi \leq   \min\{ \U(1), \ldots, \U(R)  \},
\label{eq::bounds-max-min}
\end{eqnarray} 
where the $\L(q)$'s and $\U(r)$'s are all functions of some population moments. 
For example, in Theorem \ref{thm:arbitrary}, $Q=R=2$, and $\{\L(1), \L(2)\}$ and $\{\U(1), \U(2)\}$ are the two terms in the expressions of $\widetilde{\textnormal{LB}}^{I}_t$ and $\widetilde{\textnormal{UB}}^{J}_t$, respectively.
For any $q$ and $r$, the point estimators of $\L(q)$ and $\U(r)$ are $\widehat{\L}(q)$ and $\widehat{\U}(r)$, which are asymptotically normal with means $\L(q)$ and $\U(r)$ and estimated standard errors $\widehat{\sigma}_\L(q)$ and $\widehat{\sigma}_\U(r)$.

\citet{imbens2004confidence} proposed a method to construct confidence intervals for partially identified parameters, but their method requires that the estimators for the upper and lower bounds follow a joint bivariate normal distribution asymptotically. \citet{chernozhukov2013intersection} proposed a general method, but the construction of their confidence interval is non-trivial. Fortunately, the bounds in our paper have a nice feature: the lower bounds are the maximums of some parameters, and the upper bounds are the minimums of some parameters. This feature allows us to extend \citet{imbens2004confidence}'s method to construct confidence intervals for $\pi$. 
We first review their method. They considered a simple case in which the parameter of interest has bounds $\L(q) \leq \pi \leq \U(r)$ for fixed $q$ and $r$, and proposed to use 
\begin{eqnarray}
\label{eq::old-ci}
\CI(q,r)=\left[  
\widehat{\L}(q) - C \times \widehat{\sigma}_\L(q),
\widehat{\U}(r) + C \times \widehat{\sigma}_\U(r)
\right],
\end{eqnarray} 
as a $(1-\alpha)$-level confidence interval for $\pi$, where $\Phi(\cdot)$ is the cumulative distribution function of a standard normal random variable, and $C$ is determined by
\begin{eqnarray}
\label{eq::old-ci-C}
\Phi\left[ C+ \frac{\widehat{\U}(r) -   \widehat{\L}(q)     }{\max\{ \widehat{\sigma}_{\L}(q),\widehat{\sigma}_{\U} (r) \} }\right] -\Phi(-C)=1-\alpha.
\end{eqnarray}
Note that with large samples, the solution of $C$ in equation \eqref{eq::old-ci-C} is close to $\Phi^{-1}(1-\alpha)$, the one-sided critical value based on a standard normal distribution. See \citet{vansteelandt2006ignorance} for similar discussion.

However, the bounds in \eqref{eq::bounds-max-min} have more complicated forms. If we know the true indices of the bounds $q_0 = \arg\max_{1\leq q\leq Q} \L(q)$ and $r_0 = \arg\min_{1\leq r\leq R} \U(r)$, then the bounds in \eqref{eq::bounds-max-min} are simply $\L(q_0) \leq \pi \leq \U(r_0)$, and we can use $\CI(q_0, r_0)$ as a confidence interval for $\pi.$ In our case, the true values of $q_0$ and $r_0$ are unknown, we can first obtain their estimators $\widehat{q} = \arg\max_{1\leq q\leq Q} \widehat{\L}(q)$ and $\widehat{r} = \arg\min_{1\leq r\leq R}\widehat{\U}(r)$, and then use $\CI(\widehat{q}, \widehat{r})$ as a confidence interval for $\pi$. In the following subsection, we will prove the validity of this new method for constructing confidence intervals.

Intuitively, as the large sample size goes to infinity, $\widehat{q}$ and $\widehat{r}$ will converge to the true values $q_0$ and $r_0 $, and   $\CI(\widehat{q}, \widehat{r})$ will then converge to $\CI(q_0, r_0)$. Because $\CI(q_0, r_0)$ has a correct asymptotic coverage rate as shown in  \citet{imbens2004confidence}, $\CI(\widehat{q}, \widehat{r})$ also has a correct asymptotic coverage rate for $\pi.$ The following theorem formally ensures the validity of this new confidence interval.

\begin{theorem}
\label{thm::new-ci}
If (1) $\{ \L(1), \ldots, \L(Q)\}$ have a unique maximum value $\L(q_0)$, and $\{ \U(1), \ldots, \U(R) \} $ have a unique minimum value $\U(r_0)$; (2) for any $q$ and $r$, the asymptotic distribution of $\{\widehat{ \L}(q), \widehat{\U}(r) \}$ is bivariate normal with means $\{ \L(q),\U(r)\}$ and estimated standard errors $\{ \widehat{\sigma}_{\L}(q),  \widehat{\sigma}_{\U}(r) \}$, then $\CI(\widehat{q}, \widehat{r})$ has a coverage rate at least as large as $1-\alpha$ asymptotically.
\end{theorem}

Technically, in Theorem \ref{thm::new-ci} there are two types of coverage rates (the pointwise and uniform coverage rates), and they require different regularity conditions \citep{imbens2004confidence,vansteelandt2006ignorance, romano2010inference,chernozhukov2013intersection}.  For simplicity we relegate the technical discussions to the online supplementary material. 
Intuitively, these conditions rule out the extreme cases that the $\L(q)$'s and $\U(r)$'s are too close.   In our application, the estimated bounds are wide enough for us to believe that $\L(q)$'s and $\U(r)$'s are not close even in the presence of statistical uncertainty. Therefore,  the confidence intervals calculated in our application are likely to have coverage rates at least as large as 95\% asymptotically.
According to Theorem \ref{thm::new-ci}, for each possible value of the parameter, 
$\CI(\widehat{q}, \widehat{r})$ has a coverage rate at least as large as $1-\alpha$ asymptotically, and in the case of partial identification it is inevitable that for some values of the parameter, $\CI(\widehat{q}, \widehat{r})$ will have higher coverage rates than the nominal level. 
We give a numerical example to illustrate the procedure of Theorem 3  for constructing confidence intervals.

\begin{example}
Suppose that the parameter of interest, $\pi$, has bounds
$\max\{ \L(1), \L(2) \}  \leq \pi \leq   \min\{ \U(1), \U(2)  \}$. Suppose further that, from the observed data, we have point estimates $\widehat{ \L}(1)=-0.2$, $\widehat{ \U}(1)=0.5$, $\widehat{ \L}(2)=-0.1$ and $\widehat{ \U}(2)=0.8$, with asymptotic standard errors $\widehat{\sigma}_{\L}(1)=0.01$, $\widehat{\sigma}_{\U}(1)=0.04$, $\widehat{\sigma}_{\L}(2)=0.02$ and  $\widehat{\sigma}_{\U}(2)=0.03$.  Note that these numbers are artificial, not from the motivating example.

We can construct the confidence interval of $\pi$ in the following steps.
First, we  calculate $\widehat{q}=\arg\max_{q=1,2} \L(q)=2$ and $\widehat{r}=\arg\min_{r=1,2} \U(r)=1$. Second,
we obtain $C=1.645$ by solving (6)  with $q$ and $r$ replaced by $\widehat{q} = 2$ and $\widehat{r} = 1$.
Third, we calculate $\CI(\widehat{q}, \widehat{r})   = \CI(2,1)  =[-0.133,0.566]$ according to (5), which is a confidence interval for $\pi$.
\end{example}
%


The bounds with a single time point correspond to the case with $Q=R=1$, and $\CI(\widehat{q}, \widehat{r})$ reduces to the one proposed by \citet{imbens2004confidence}. The bounds with multiple time points correspond to the case with $Q=R=2$, and we can use $\CI(\widehat{q}, \widehat{r})$ to construct a confidence interval for parameter $\pi_t$ in our missing data problem.  

In the  supplementary material, we conduct simulation studies to evaluate the method for constructing confidence intervals. The simulation studies show that the coverage rates of the confidence intervals are close to the nominal levels in a wide range of situations with moderate sample sizes.


Although it is not the focus of our main paper, in the supplementary material, we also consider constructing confidence intervals for the bounds themselves. Because of the coverage guarantees for the bounds, the method in the supplementary material can be used to construct confidence intervals for the parameter of interest when condition (1) in Theorem \ref{thm::new-ci} fails.

\subsection{Extension to the case adjusting for discrete covariates}
In our data, there are some discrete covariates such as gender and region, which are prognostic to the HIV status. We also find that these two covariates are related to the missing data indicator. It is therefore more reasonable to impose Assumption \ref{thm:single} conditioning on these covariates. In this subsection, we show how to use our method to construct confidence intervals with discrete covariates.
In practice, covariates help to improve inference in three ways. First, the assumptions will generally be more plausible conditional on covariates. Second, we can first calculate the bounds conditional on the covariates and then average over them to obtain tighter bounds of the whole population \citep{lee2009training,long2013sharpening,mealli2013using}. Third, covariates help to improve the estimation precision.

Consider a discrete covariate $X$ corresponding to $K$ subpopulations, with $w_k = \pr(X=k)$ being the proportion of subpopulation $k.$ We fix the $w_k$'s at the sample frequencies of $X$, and treat them as known constants. Therefore, our inference is conditional on the proportions of the subpopulations. The parameter of interest in subpopulation $k$, $\pi_k$, has bounds of the following forms
\begin{eqnarray}
\max\{ \L_k(1), \ldots, \L_k(Q_k) \}  \leq \pi_k \leq   \min\{ \U_k(1), \ldots, \U_k(R_k)  \}.
\label{eq::bounds-max-min:cov}
\end{eqnarray} 
where the subscript $k$ is the index for the quantities of the $k$-th subpopulation analogues of those in Section \ref{subsec::new-ci1}. We are interested in the overall bounds of $\pi=\sum_{k=1}^Kw_k\pi_k.$ We first consider the following simple form of bounds with known and fixed $q_k$'s and $r_k$'s:
$$
\sum_{k=1}^Kw_k\L_k(q_k) \leq \pi \leq \sum_{k=1}^Kw_k\U_k(r_k).
$$
If the joint distribution of the estimators of the upper and lower bounds are asymptotically normal, then we can again construct a confidence interval for $\pi$
using \citet{imbens2004confidence}'s method, denoted by $\CI(q_1,\ldots,q_k,r_1,\ldots,r_k)$. 
Let the true indices for the $k$-th subpopulation bounds be $q_{k0} = \arg\max_{1\leq q\leq Q_k} \L_k(q)$ and $r_{k0} = \arg\min_{1\leq r\leq R_k} \U_k(r)$. They are unknown, but can be consistently estimated by the sample analogues $\widehat{q}_{k0} = \arg\max_{1\leq q\leq Q_k} \widehat{\L}_k(q)$ and $\widehat{r}_{k0} = \arg\min_{1\leq r\leq R_k} \widehat{\U}_k(r)$. We then construct the final confidence interval for $\pi$ as
$\CI(\widehat{q}_1,\ldots,\widehat{q}_k,\widehat{r}_1,\ldots,\widehat{r}_k)$. 
The following corollary extends Theorem \ref{thm::new-ci}, justifying the above confidence interval with a discrete covariate.

\begin{corollary}
\label{cor::new-ci:cov}
If (1) for all $k$, $\{ \L_k(1), \ldots, \L_k(Q_k)\}$ have a unique maximum value, and $\{ \U_k(1), \ldots, \U_k(R_k) \} $ have a unique minimum value; (2) for any $q_k$'s and $r_k$'s, the asymptotic distribution of $\left\{\sum_{k=1}^Kw_k\widehat{ \L}_k(q_k)\right.$,\\
 $\left.\sum_{k=1}^Kw_k\widehat{\U}_k(r_k) \right\}$ is bivariate normal, then  $\CI(\widehat{q}_1,\ldots,\widehat{q}_k,\widehat{r}_1,\ldots,\widehat{r}_k)$
has a coverage rate at least as large as $1-\alpha$ asymptotically.
\end{corollary}

For continuous covariates, we need to assume parametric or semiparametric models, and we leave this topic for future research. For time-varying covariates, we can treat them as covariates at different time points, and obtain the bounds and confidence intervals with the same procedure.

\section{Application}\label{sec::app}

\subsection{Background and motivation}
Credible estimates of the prevalence of HIV are essential for policy makers to plan control programs and interventions. However, population-based surveys may be affected by  missing data on respondents' HIV status \citep{arpino2014using}. 
The Malawi Diffusion and Ideational Change Project is a longitudinal survey conducted in rural Malawi every two years since 1998. 
The survey is from a collaboration between the University of Pennsylvania and the College of Medicine and Chancellor College at the University of Malawi. The data can be downloaded from \url{ http://www.malawi.pop.upenn.edu} and include the outcomes of HIV tests for the years 2004, 2006 and 2008.
We will give a data description that is relevant to our context, and refer to the original study \citep{anglewicz2009malawi} for more details.
The project started with a main survey, which collected information on household structure, sexual relations, marriage and partnership histories, etc. Starting from 2004, the Voluntary Consulting and Test survey was added to the main survey, which consisted of a short questionnaire and free tests for HIV and other sexually transmitted infections. 

However, the HIV status is missing for a substantial fraction of the sample.  There are different types of missing data, including temporary absence, loss of results, relocation, hospitalization and refusal to participate. 
In the analysis of \citet{arpino2014using}, they ignored different types of missing data, and used the binary missing data indicator  to obtain the bounds of the HIV prevalence. With cross-sectional data, they calculate the worst-case bounds; with longitudinal data, they improve their bounds under monotonicity.  Unfortunately, the bounds they obtained are quite wide, e.g., the bound of the HIV prevalence in the south region in 2008 is [0.035, 0.529] with cross-sectional data and [0.082,0.529] with longitudinal data.
In the following analysis, we are aiming to improve the inference by taking into account the difference in missing types.

\subsection{Data description}

Following the analysis of \citet{arpino2014using}, we focus on people who were interviewed in 2004, and drop units who were never successfully contacted.  We use the data in years 2004, 2006 and 2008. Because HIV prevalence is defined for the population of alive individuals, we consider only the alive people when computing HIV prevalences for each of the years 2004, 2006 and 2008.
The data are available from the online materials of \citet{arpino2014using}. 
The total sample size is 4062.
The HIV status is missing for 1185, 1487  and 1706 individuals in 2004, 2006 and 2008, with missing proportions 29.2\%, 36.6\% and 42.0\%, respectively.
In the data, the reasons for the missing data in the main survey are categorized as ``refused,'' ``hospitalized,'' ``not known,'' ``temporarily absent,'' ``moved.'' Other reasons are classified as the residual category ``other,'' consisting mainly of people who did not fill in the questionnaire because they were too old or too sick, or for unspecified reasons that may also include migration \citep{arpino2014using}.
 The reasons for the missing data in the Voluntary Consulting and Test survey are categorized as ``refused,'' ``hospitalized,''  ``temporarily absent,''  ``not known,'' ``moved'' and   ``results lost.'' Other reasons for item non-response are classified  as the category ``other.''
  The HIV status is missing if either data from the main survey or the Voluntary Consulting and Test survey are missing.
 In our analysis, we define $R_t=-1$ if the reason of missingness in either of the two surveys belongs to the categories ``refused,'' ``hospitalized'' or ``others.''
 For the rest of the  reasons of missingness, we define $R_t=0$. Table~\ref{tab:missing} summarizes the distributions of the missing types.

     In our data, there may be some reasons that are related to the HIV status but we classify them as missing at random. To deal with this, we will propose an approach to conducting sensitivity analysis in Section 5.4.
 We will not consider measurement error, because as stated in \citet{arpino2014using}, measurement error in the two types of test (oral swabs and blood test) appears to be small.

\begin{table}
\begin{center}
\caption{Numbers  of missing types by year with proportions in parentheses. The sample sizes across years differ because some units died before observations.}
\label{tab:missing}
\begin{tabular}{cccc} \hline
Year & $R=-1$ & $R=0$ & $R=1$ \\ \hline
2004 & 386 (9.5\%)& 799 (19.7\%)  & 2877 (70.8\%) \\
2006 & 323 (8.0\%)  &1164 (29.0\%) & 2531 (63.0\%) \\
2008 & 453 (11.5\%)& 1253 (31.8\%) &2233 (56.7\%) \\  \hline
\end{tabular}
\end{center}

\end{table}

\subsection{Data analysis}

The survey was carried out in three administrative regions:  center, south and north. Because these regions have very different demographic characteristics, we conduct  analyses within subpopulations stratified by region and gender. Within each level of region and gender, it is plausible that the absence of the individual or the loss of the individual's result does not depend on the individual's HIV status, i.e., Assumption \ref{asm:pmar} holds. Because of the feature of HIV infection discussed in Section 3.1, Assumption \ref{asm:strmon} holds. Therefore, we can apply Theorems \ref{thm:single} and \ref{thm:arbitrary} to calculate the bounds of HIV prevalence.
 The upper and lower bounds have explicit forms, and we can estimate the bounds by replacing the probability parameters by their sample frequency analogues, and then apply the method in Section \ref{sec::new-ci} to construct confidence intervals.
We empirically check the testable conditions for all three years, and find that the conditions hold in all subpopulations and years. Therefore, the data provide no evidence against Assumption \ref{asm:pmar}, because Assumption \ref{asm:strmon} holds automatically.

\begin{figure}
  \begin{minipage}[t]{1\linewidth}
	    \centering
	    \includegraphics[width=\textwidth]{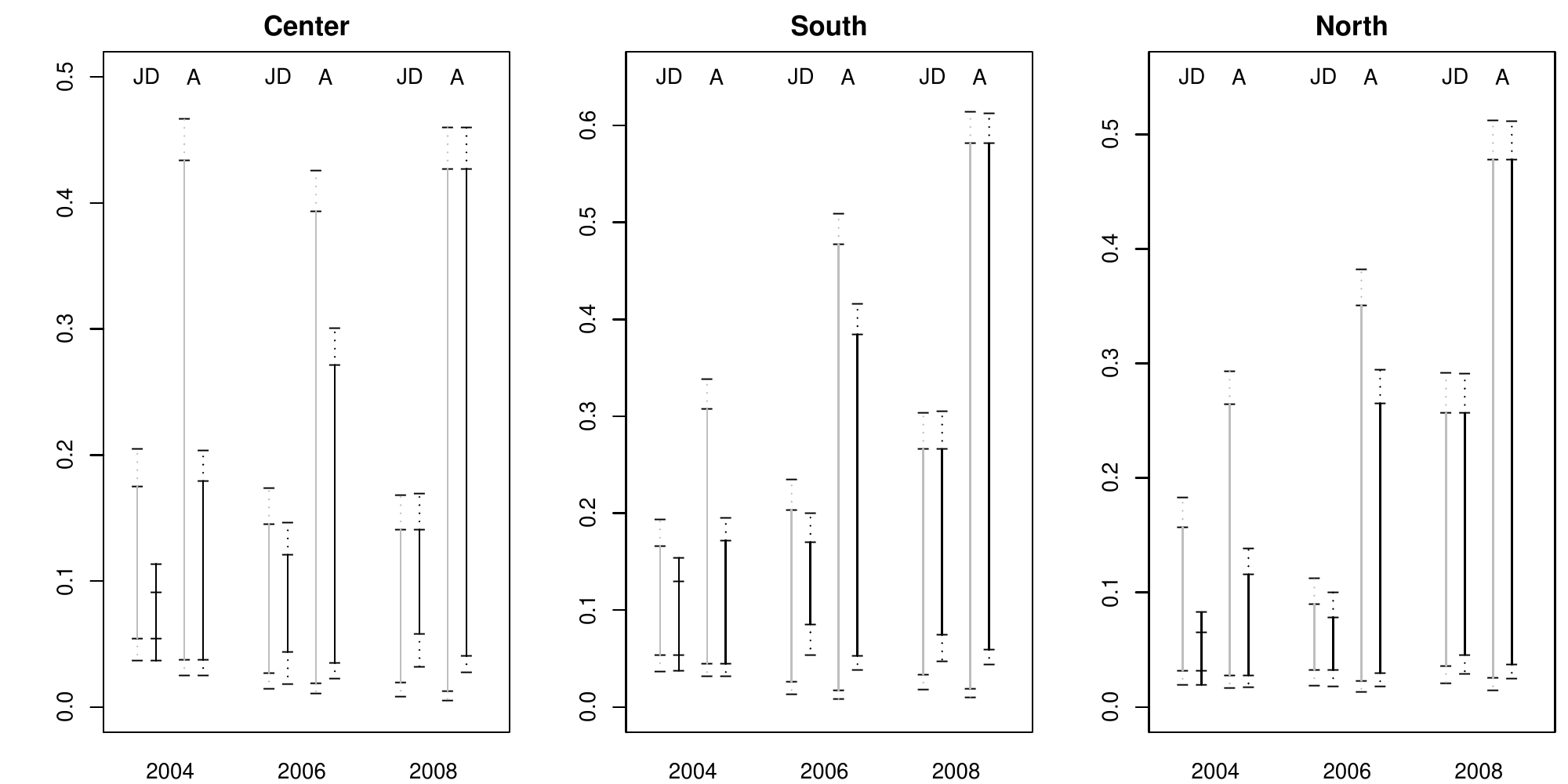}
	    \captionsetup{labelformat=empty}
       \caption*{(a) Men in the three administrative regions.}
	    \label{fig::arpino1}
	  \end{minipage}
	  \\
	  \begin{minipage}[t]{1\linewidth}
	    \centering
	    \includegraphics[width=\textwidth]{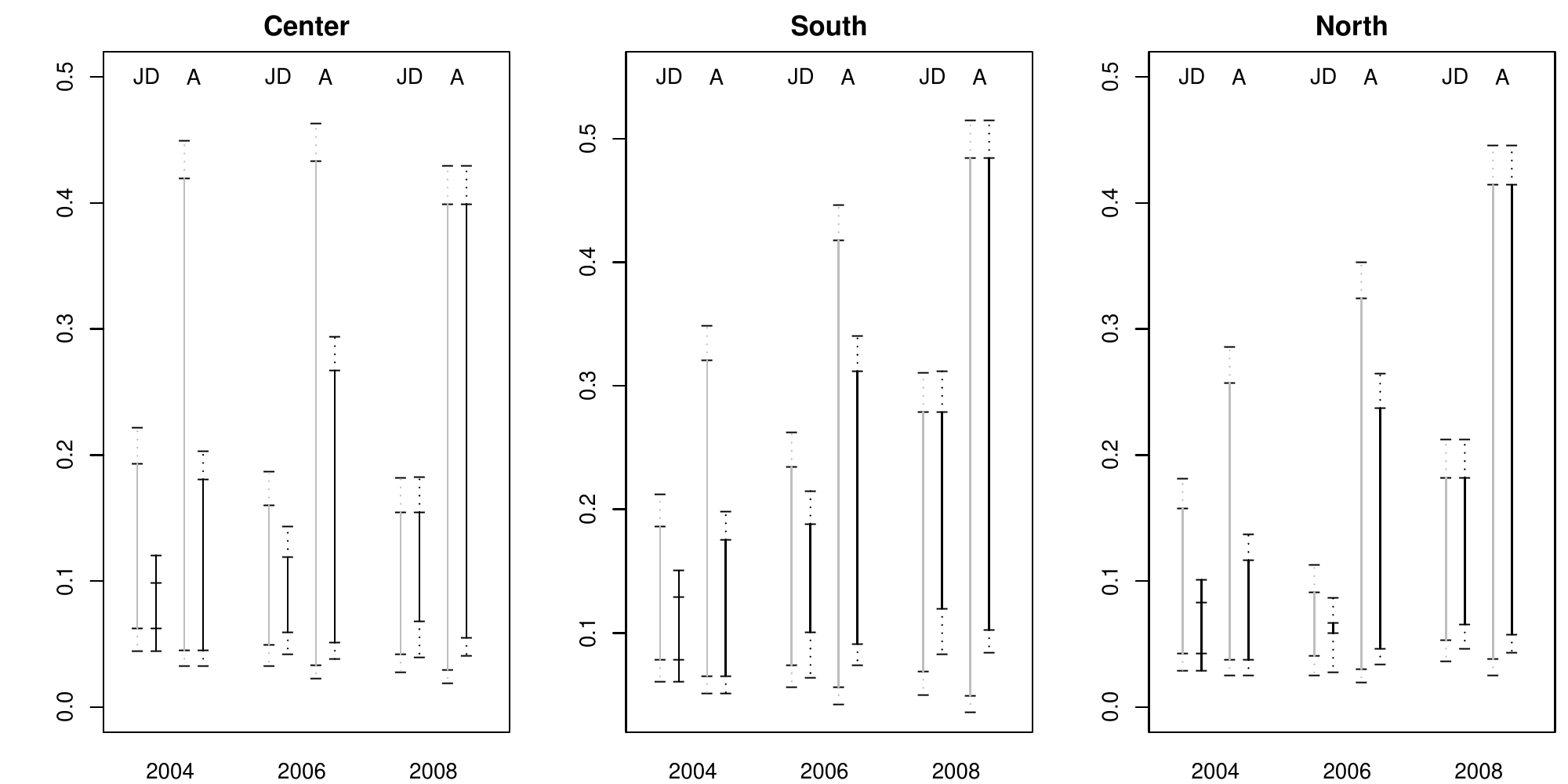}
	     \captionsetup{labelformat=empty}
	    \caption*{(b) Women in the three administrative regions.}
	    \label{fig::arpino2}
	  \end{minipage}
 \caption{The solid lines are the bounds and the dotted extended lines are the 95\% confidence intervals  for the HIV prevalence in the corresponding years. The grey lines are the results with a single time point and the black lines are the results with multiple time points. The label ``JD'' corresponds to the result of our method, and the label ``A'' corresponds to the results of \citet{arpino2014using}'s method. }
  \label{fig::arpino} 
\end{figure}


Figure \ref{fig::arpino}  shows the estimated bounds and confidence intervals for the HIV prevalence from our method and \citet{arpino2014using}'s method for the three administrative regions of the country separated by gender. For descriptive convenience, we call \citet{arpino2014using}'s method the ``A method,'' and our method the ``JD method'' from now on.
For women in the north region in 2006 (the third plot of Figure \ref{fig::arpino}(b)), the  A method gives bounds $(0.047,0.237)$, and the  JD method gives bounds $(0.058,0.067)$; the  JD method reduces  95.3\% of the width of the bounds.  For men in the south region in 2004 (the second plot of Figure \ref{fig::arpino}(a)),  the  A method gives bounds  $(0.045,0.172)$, and the JD method gives bounds $(0.054,0.129)$; the JD method reduces 40.6\% of the width of the bounds. The reductions in the widths of the bounds  in these two subpopulations are the largest and smallest among all subpopulations, respectively.
For men in the north region in 2006 (the third plot of Figure \ref{fig::arpino}(a)), the A method gives a confidence interval $(0.013,0.382)$, and the JD method gives a confidence interval $(0.019,0.113)$; the JD method reduces 75.2\% of the width of the confidence interval.
For men in the south region in 2004 (the second plot of Figure \ref{fig::arpino}(a)), the  A method gives a confidence interval $(0.032,0.195)$, and the  JD method gives a confidence interval $(0.037,0.154)$; the JD method reduces 28.5\% of the width of the confidence interval. The reductions in the widths of the confidence intervals  in these two subpopulations are the largest and smallest among all subpopulations, respectively. 

\begin{figure}
\begin{center}
\includegraphics[width=\textwidth]{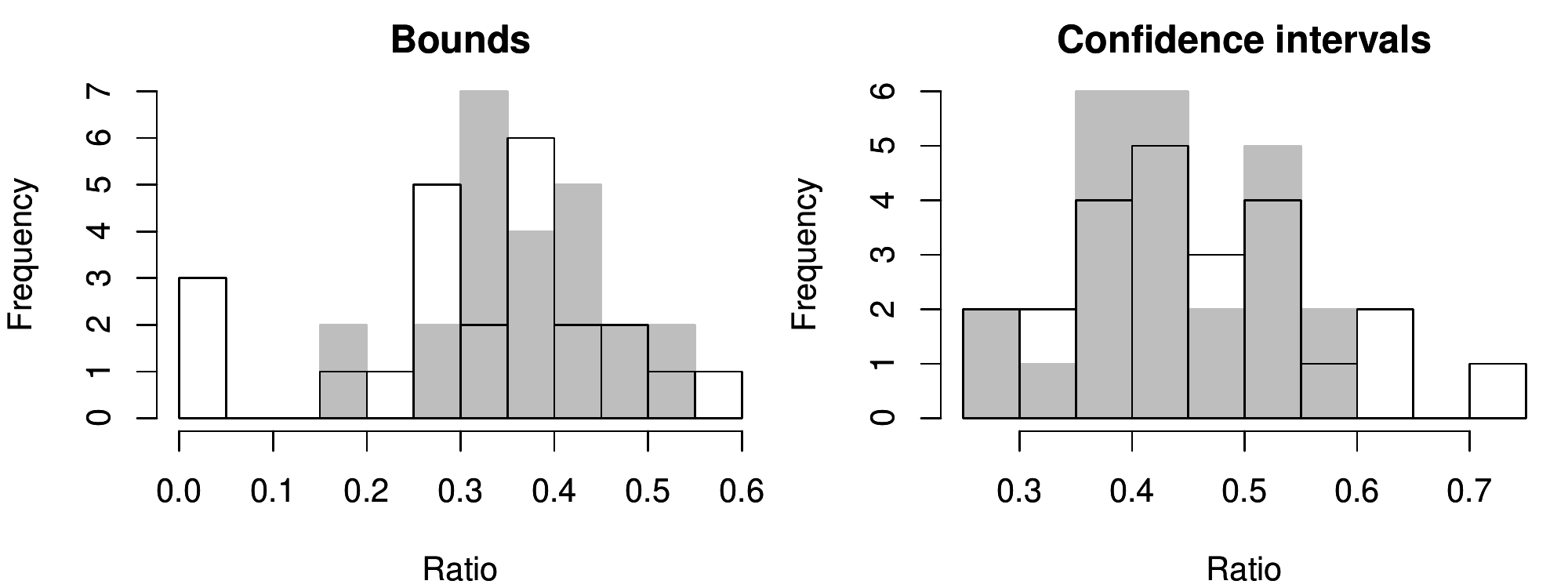}
\end{center}
\caption{Ratios  of the widths of the bounds and confidence intervals for the HIV prevalence obtained by the JD and A methods. The grey histograms are the results with a single time point and the white histograms are the results with multiple time points.}
\label{fig:ratio}
\end{figure}

Figure \ref{fig:ratio} shows the histograms of the ratios  of the widths of the bounds and confidence intervals obtained by the  JD method divided by those obtained by the  A method. For all the results with a single time point and multiple time points, the ratios of the widths of the bounds range from $0.043$ to $0.594$, and the ratios of the  widths of the confidence intervals range from $0.255$ to $0.715$.
In most subpopulations, the reductions in widths of the bounds and confidence intervals are larger than 50\%, demonstrating substantial improvement in inference by taking into account  the different types of missing data.


\begin{figure}
\begin{center}
\includegraphics[width=\textwidth]{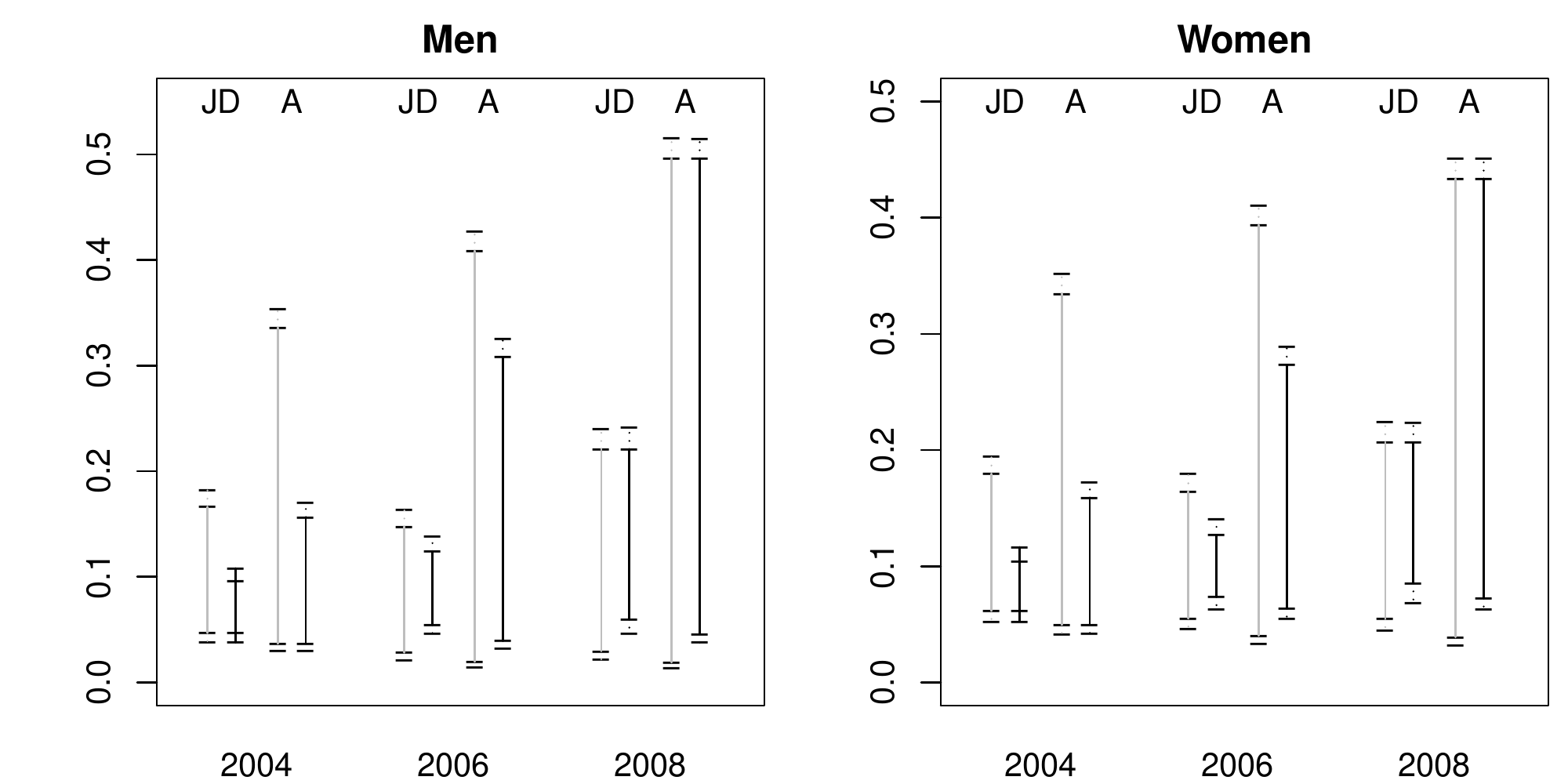}
\end{center}
\caption{The solid lines are the bounds and the dotted extended lines are the 95\% confidence intervals for the HIV prevalence in the corresponding years. The grey lines are the results with a single time point and the black lines are the results with multiple time points. The label ``JD'' corresponds to the result of our method, and the label ``A'' corresponds to the results of \citet{arpino2014using}'s method.}
\label{fig::arpino:overall}
\end{figure}

We can obtain the overall bounds of men and women by first calculating the bounds in each region and then averaging the bounds over regions. Figure \ref{fig::arpino:overall} shows the overall bounds and corresponding confidence intervals using the  A and  JD methods. The reductions in widths of the bounds and confidence intervals are large by the JD method.

 According to Figure \ref{fig::arpino}, in years 2006 and 2008, the lower bound of  the HIV prevalence in Southern Malawi is the highest among the three regions for both men and women. But in 2004, the central region has the highest lower bound. This is similar for the upper bound. However, because the bounds and confidence intervals overlap, we make no attempt to draw inference about the differences in HIV prevalence  across different regions. In Figure \ref{fig::arpino:overall}, the lower bounds and upper bounds in 2008 are the highest among the three years for both men and women, suggesting an increasing trend of the HIV prevalence.

Note that \citet{arpino2014using} used some variables as instrumental variables or monotone instrumental variables to further narrow their bounds \citep{manski2003partial, manski2000monotone}, which, however, invoked additional assumptions. As a future work, we can further narrow our bounds by combining with some of these instrumental variable assumptions.

\subsection{Sensitivity analysis}
The classification of the reasons of missingness requires prior knowledge, and sometimes it is hard to justify whether a specific reason is missing at random or not at random. For example, in our application, it is possible that absence of the individual is related to HIV status in various ways. As a result, the reason ``temporarily absent'' may be missing not at random, but in previous analysis, we define it as $R_t=0$. 
In this case, we can conduct a sensitivity analysis by gradually increasing or decreasing the set of reasons that are missing at random. Specifically, we start with the analysis that assumes all the reasons are missing not at random. Then, we change the categories of the reasons of missingness from $R_t=-1$ to $R_t=0$ one by one.  As a result, we can see how the bounds and confidence intervals change as we strengthen the assumption about the missing types.

\begin{figure}
\begin{center}
\includegraphics[width=\textwidth]{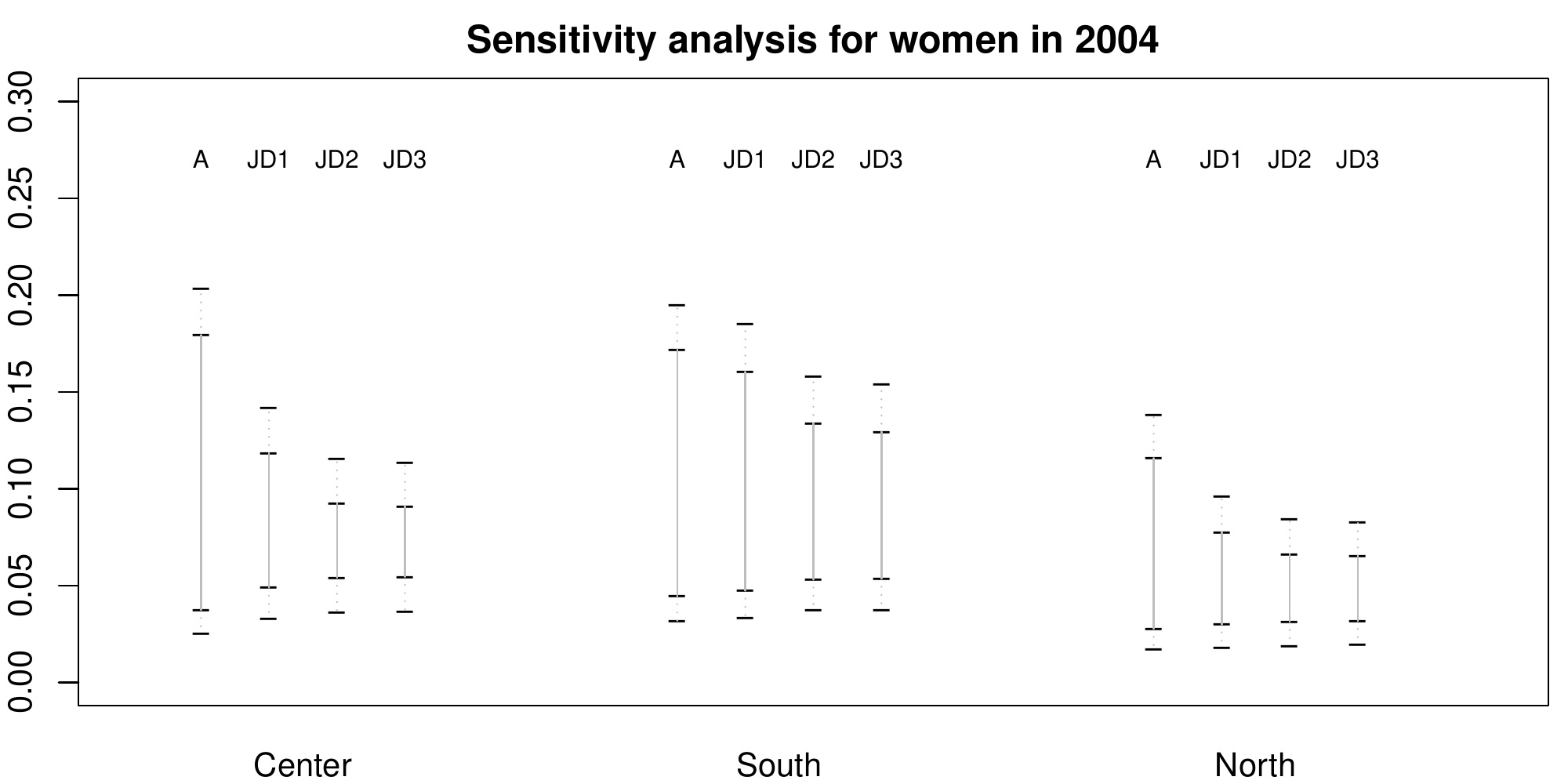}
\end{center}
\caption{The solid lines are the bounds and the dotted extended lines are the 95\% confidence intervals. The label ``A'' corresponds to the results of \citet{arpino2014using}'s method. The label ``JD1'' corresponds to the result of our method with $R_t=0$ indicating ``moved.'' The label ``JD2'' corresponds  to the result of our method with $R_t=0$ indicating ``moved'' and ``results lost or not known''. The label ``JD3'' corresponds to the result of our method with $R_t=0$ indicating ``moved,'' ``results lost or not known'' and ``temporarily absent'', which is the same as what we did in Section 5.3.}
\label{fig::arpino:sens}
\end{figure}

We conduct the sensitivity analysis for the bounds with multiple time points in 2004 in the subpopulation of women. We start with the method in \citet{arpino2014using}, which classifies all the reasons of missingness as $R_t=-1$.  
We then  change the categories of    reasons ``moved,''  ``results lost or not known,''  and ``temporarily absent'' to $R_t=0$  one by one.  
The results in Figure \ref{fig::arpino:sens} show that even with a single reason classified as missing at random, we have meaningful improvement, and with three reasons classified as missing at random, we have substantial improvement. Analogously, we can conduct sensitivity analyses for other subpopulations, but we omit the results to save space.

 \section{Discussion}\label{sec::discussion}
 
%

We discussed binary outcomes in Sections \ref{sec::single} and \ref{sec::longitudinal}. For general outcomes, we can dichotomize the outcomes, and apply our results for binary outcomes to obtain bounds on the distribution functions. Furthermore, bounds on the distribution function can be used to construct bounds on quantiles as suggested by \citet{manski2009identification}. 

In the application, there are many types of missing data. We collapsed them into two categories: one  is more likely to be missing at random, and the other  tends to be related to the missing outcome itself. By doing this, we do not need to model the detailed missing data patterns. Ideally, we could further distinguish the detailed missing types. But this will require more assumptions about the missing data mechanism. From the HIV data, although we know there are different missing types, we do not have very clear understanding about the missing data mechanism. 
If we are willing to impose  parametric models, it is promising that more efficient estimates could be obtained. Our paper, however, focuses more on nonparametric and robust analyses. 

The current bounds are sharp, that is, without imposing further assumptions beyond Assumptions \ref{asm:pmar} and \ref{asm:strmon}, we cannot improve these bounds. In particular, we do not impose any assumptions about the transition probabilities of the missing indicator $R_t$. If we are willing to model the transition probability matrix, then we can obtain narrower bounds. However, from the HIV study, it does not seem obvious how to impose assumptions on this transition probability matrix. Therefore, we do not explore this direction in our current paper, but leave it to future research.

%

From the perspective of study designs, our results suggest that in addition to recording the binary missing indicator, researchers should also collect nonresponse types and, if possible, distinguish between types related to the outcomes and types unrelated to the outcomes.

%
%
%
%
%


\section*{Acknowledgements}

Drs. Avi Feller and Shu Yang edited early versions of the paper. Professors Peter Bickel and Eric
Tchetgen Tchetgen gave us insightful comments on pointwise and uniform coverage properties
of confidence intervals. We thank the Editor, Associate Editor and a reviewer for careful reading and many constructive comments.

\begin{supplement}
\sname{Supplement Material}\label{suppA}
\slink[url]{https://projecteuclid.org/euclid.aoas/1536652976\#supplemental}
\sdescription{The supplementary material consists of four parts. Section S1 gives the proofs of the theorems of the bounds. Section S2 gives the testable conditions with multiple time points. Section S3 gives the proofs of the theorem and corollary for constructing confidence interval. Section S4 shows the results of the simulation studies. 
}
\end{supplement}

\bibliographystyle{imsart-nameyear}
 \bibliography{partialMAR_refs}

%
%
%
%
%
%


\end{document}